\documentclass[twocolumn]{revtex4-1}
\usepackage{amsmath,amsfonts,amssymb}
\usepackage{wrapfig}
\usepackage{graphicx}
\usepackage{bbm}
\usepackage{graphics}
\usepackage{float}
\usepackage{color}

\begin{document}


\title{Frequency stabilization of the zero-phonon line of a quantum dot via phonon-assisted active feedback} 



\author{Jack Hansom}
\author{Carsten H. H. Schulte}
\author{Clemens Matthiesen}
\author{Megan Stanley}
\author{Mete Atat\"ure}

\affiliation{Cavendish Laboratory, University of Cambridge, JJ Thomson Avenue, Cambridge CB3 0HE, United Kingdom}


\date{\today}

\begin{abstract}
We report on the feedback stabilization of the zero-phonon emission frequency of a single InAs quantum dot. The spectral separation of the phonon-assisted component of the resonance fluorescence provides a probe of the detuning between the zero-phonon transition and the resonant driving laser. Using this probe in combination with active feedback, we stabilize the zero-phonon transition frequency against environmental fluctuations. This protocol reduces the zero-phonon fluorescence intensity noise by a factor of 22 by correcting for environmental noise with a bandwidth of 191 Hz, limited by the experimental collection efficiency. The associated sub-Hz fluctuations in the zero-phonon central frequency are reduced by a factor of 7. This technique provides a means of stabilizing the quantum dot emission frequency without requiring access to the zero-phonon emission.  
\end{abstract}


\maketitle
\begin{figure}
\includegraphics[width=3.37in]{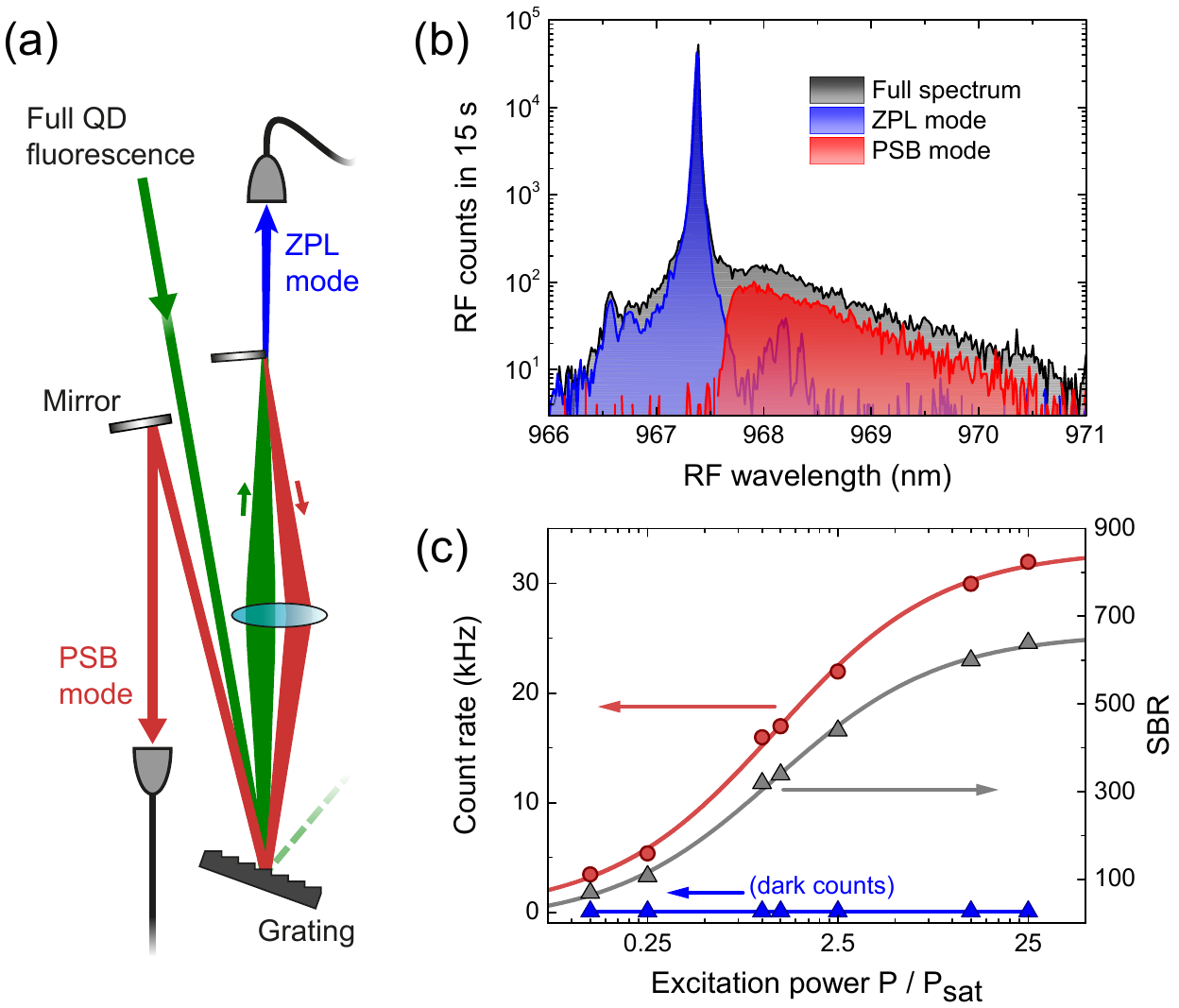}
\caption{
\label{Figure1}
(a) Schematic diagram of the apparatus used for spectral discrimination of the zero-phonon line (ZPL) from the red-detuned phonon-assisted RF. (b) RF spectra recorded from the different detection modes shown in a). The black curve shows the full emission spectrum of quantum dot RF measured by replacing the grating with a mirror. The blue (red) curve shows a spectrum with the same integration time using the ZPL (PSB) detection mode.  (c) Photon detection rates (red curve) in the PSB mode as a function of excitation power, with the corresponding signal to laser background ratio labelled as SBR (grey curve). The blue curve shows the count rates when the QD is tuned off resonance and is limited to APD dark counts ($50$ Hz). 
}
\end{figure}

\begin{figure}
\includegraphics[width=3.37in]{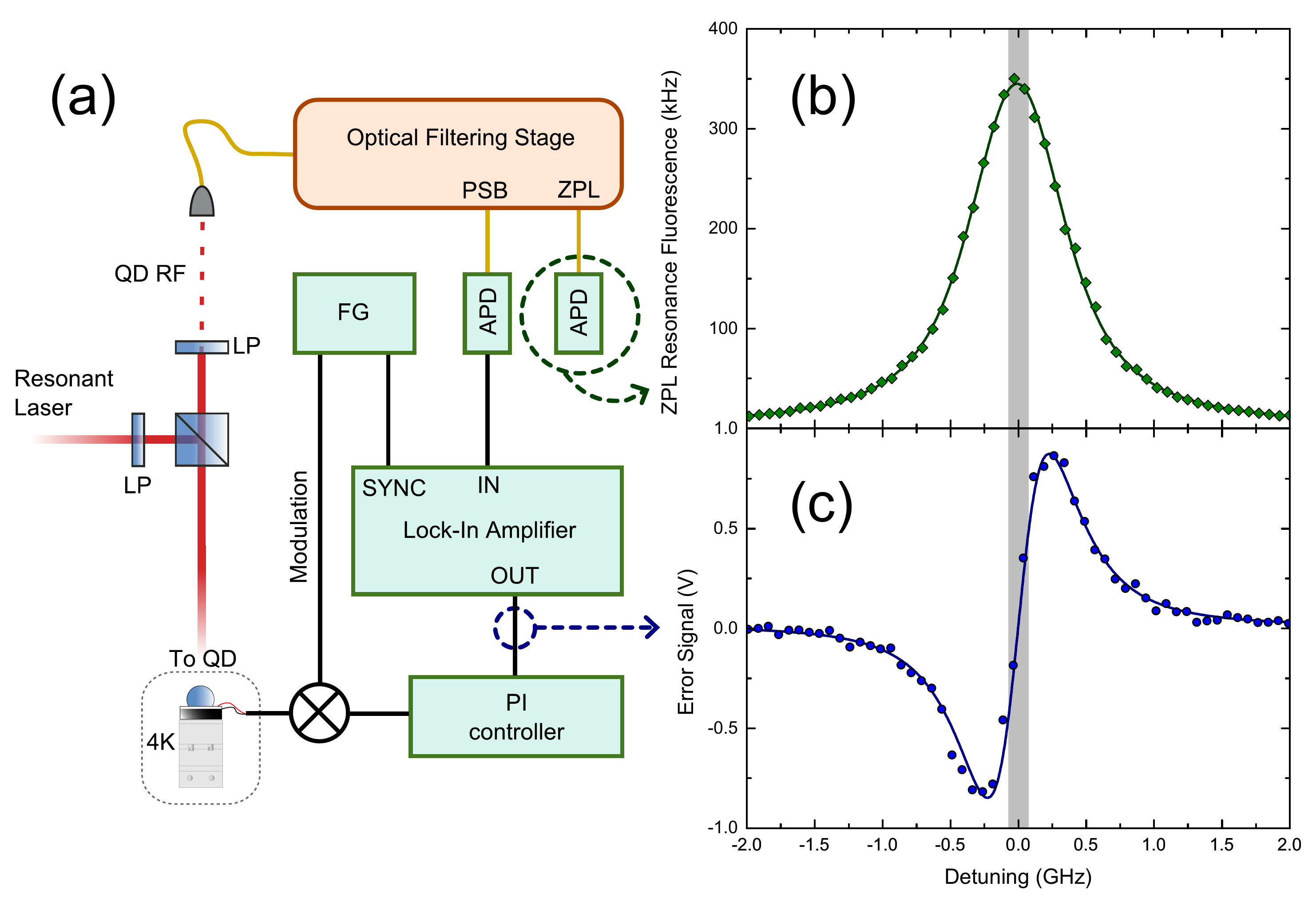}
\caption{
\label{Figure2}
(a) Schematic diagram of the setup used for frequency stabilization of the ZPL fluorescence. LP: linear polariser, APD: Avalanche PhotoDiode, FG: Function Generator. (b) Detuning dependence of QD fluorescence with a sub-linewidth gate modulation applied. The measured absorption linewidth without modulation is $610 \pm 20$ MHz, whereas the linewidth shown here is $758 \pm 6$ MHz. (c) Error signal obtained at the output of the lock-in amplifier as the DC offset of the FG is tuned across resonance. The gate modulation used here has an amplitude corresponding to the width of the grey bar and a modulation frequency of $1.5$ kHz.}
\end{figure}

\begin{figure}
\includegraphics[width=3.37in]{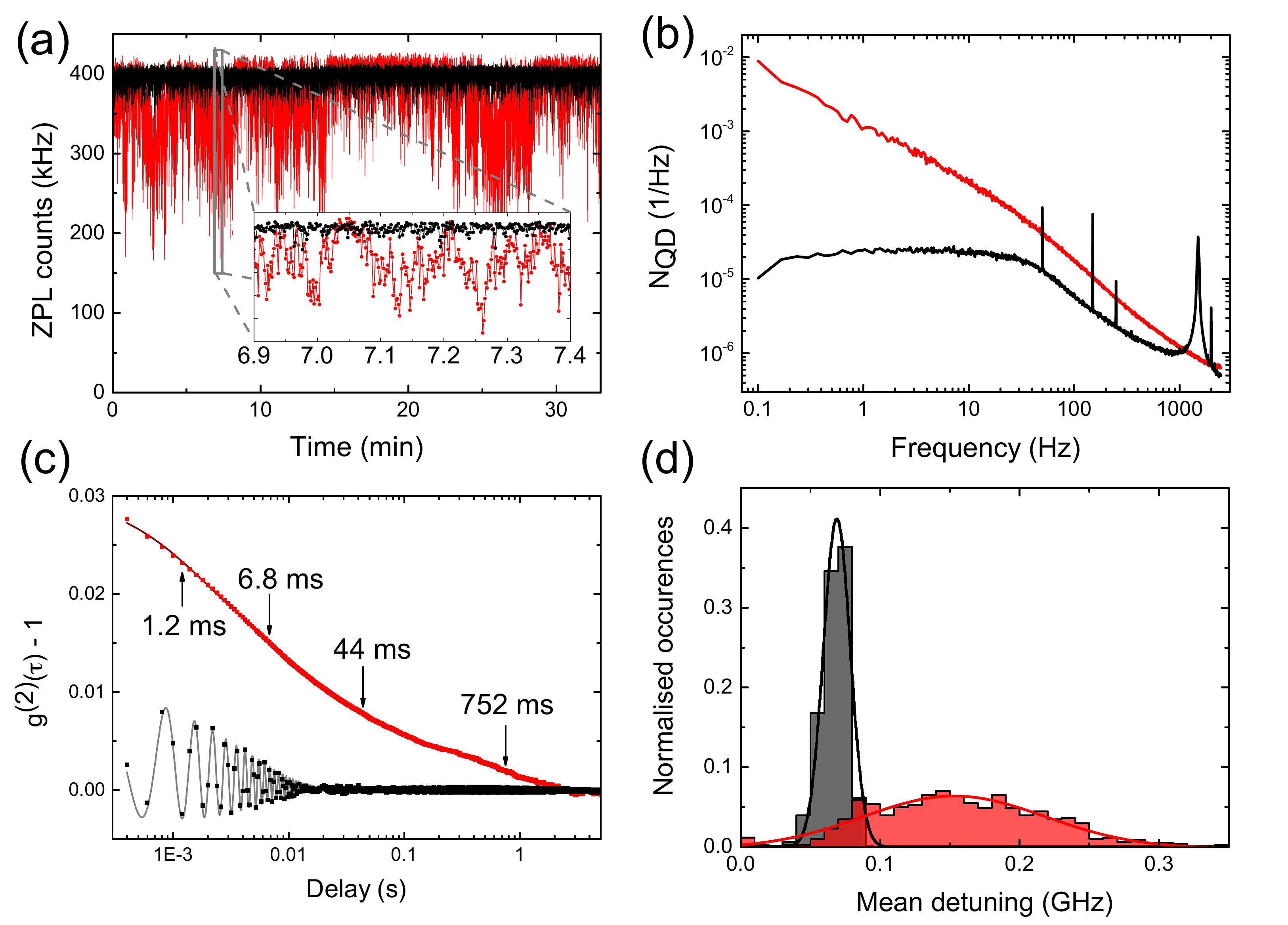}
\caption{
\label{Figure3}
(a) Time traces of the ZPL-RF count rates with (black curve) and without (red curve) frequency stabilization at $100$ ms integration time. The inset shows a zoom-in. (b) ZPL RF noise power spectrum with (without) stabilization shown in black (red). A laser background measurement at similar count rates (not shown) was subtracted from both curves. (c) Autocorrelations calculated from the same time traces as (b). The black (red) data points show the autocorrelation with (without) the stabilization scheme. The red curve shows the fit to exponential decays with $4$ characteristic timescales as indicated with arrows. The gray curve is a fit to four exponential decays with timescales fixed by the red fit, with an exponentially decaying $1.5$ kHz oscillation added to fit the gate modulation and feedback response time. (d) Distribution of mean ZPL frequency detuning from stabilized laser frequency extracted from histogram analysis of $1$ s long time traces with (without) stabilization shown in black (red). The frequency fluctuations are quantified by the width of this distribution which reduces from $154$ MHz to $23$ MHz with the application of the stabilization scheme.
}
\end{figure}

 
%
%
A robust single photon source is an integral component for the implementation of many quantum technologies including linear optical quantum computing\cite{Knill2001,Kok2007}, quantum relays\cite{Froehlich2013}, and quantum networks\cite{Kimble2008}. Indium Arsenide self-assembled quantum dots (QDs) offer one of the most promising platforms\cite{Buckley2012} for such applications due to the large tuneability of their emission frequency\cite{Trotta2012}, fast triggered emission rates\cite{Bennett2005,Hargart2013}, and the possibility of integration into nanostructures\cite{Lodahl2013}. However, charge carriers confined to a semiconductor QD interact with the solid-state environment, in which disparate noise sources cause fluctuations of the central frequency of the optical transitions\cite{Kuhlmann2013,Stanley2014,Matthiesen2014}. These fluctuations lead to both spectral distinguishability of the fluorescence from independent QDs as well as a decreased emission intensity in resonant excitation and represent a challenge to be overcome in order to achieve a scalable architecture using QDs.  Here, we develop an active feedback stabilization scheme which makes use of the phonon-assisted fluorescence intensity as a probe for fluctuations of the zero-phonon line (ZPL) frequency. Our scheme allows for the detection and correction of environment-induced frequency fluctuations in a manner compatible with broadband photon collection strategies.  \\
Ideally, the frequency of the QD transition is stabilized without sacrificing any of the photon emission with a high bandwidth and in resonant excitation. Indeed, proposals for efficient linear optical computing with a polarization-entangled photon source require photon collection and detection efficiencies above 0.5\cite{Varnava2008,Gong2010}. Previously demonstrated frequency stabilization protocols with solid state emitters have a bandwidth limited by periodic measurements of ZPL peak position\cite{Acosta2012,Akopian2013}, or rely on continuously detecting a fraction of the ZPL\cite{Prechtel2013}. In this letter, we present a technique for simultaneous and discriminatory detection of both the zero-phonon and the phonon-assisted components of the resonance fluorescence (RF) from a single QD. We use the latter to generate an error signal in order to stabilize the intensity of the ZPL emission with a high bandwidth. We show that the stabilization scheme leads to both a broadband decrease in ZPL intensity noise, as well as a reduction in the underlying inhomogeneous broadening of the transition. \\
Due to the coupling of confined excitons to acoustic phonon modes, the emission spectrum of a QD exhibits a broad phonon sideband (PSB) close to the ZPL\cite{Besombes2001,Favero2003}. The presence of this PSB is an additional source of spectral distinguishability for subsequently emitted photons. However, the coupling strength for low energy acoustic phonon modes tends to zero resulting in an unbroadened ZPL\cite{Krummheuer2002}, which can be used for high-visibility photonic interference experiments after filtering the PSB\cite{Matthiesen2013,He2013,Gao2013,Gazzano2013}. At a temperature of $4.2$ K the occupation probability of the relevant longitudinal-acoustic (LA) phonon modes is low leading to phonon-assisted emission predominantly on the red side of the ZPL with a detuning of order 1 nm. In order to isolate the ZPL while also achieving a broadband collection of the PSB, we use the setup shown in Fig.\ref{Figure1}a. We resonantly excite a negative trion transition in a single QD with a Rabi frequency $\Omega=  \frac{\Gamma}{\sqrt{2}}$ , where $\Gamma$ is the radiative decay rate in angular frequency. The laser intensity (frequency) is stabilized to $\pm 1\%$ ($\pm 5$ MHz). We use a confocal microscope to collect the QD RF and extinguish the reflected laser with a cross-polarization scheme\cite{Matthiesen2012}. The RF is dispersed on a $1600$ grooves/mm diffraction grating. A mirror edge is placed in the Fourier plane on the red side of the optical axis and aligned at a small angle to normal incidence. Figure \ref{Figure1}b shows emission spectra of the ZPL (PSB) detection mode as a blue (red) curve, as well as a reference spectrum (black curve) measured by replacing the grating with a mirror. The ZPL mode spectrum shows a narrow line with a peak count rate $85\%$ that of the reference measurement, which is limited by the efficiency of the grating and the additional optics. The PSB mode spectrum shows a broadband collection of phonon-assisted fluorescence with negligible contribution from the ZPL. The separation of the two photonic modes has a twofold benefit: the ZPL mode is spectrally filtered by the grating, while the broadband collection of the PSB mode provides an additional channel for measurement of the spectral wandering of the ZPL. Figure \ref{Figure1}c shows the count rates collected in the PSB mode detected on an avalanche photodiode (APD). Count rates of $\gtrsim30$ kHz are obtained with a maximum signal-to-background ratio of $640$ limited by the detector dark count rate of $50$ Hz. By comparison, count rates of $\gtrsim700$ kHz are obtained in the ZPL mode corresponding to an overall collection efficiency of $\approx0.5\%$. The count rates and high signal-to-background ratio obtained in the PSB mode are sufficient to provide an error signal for feedback stabilization of the ZPL resonance frequency. \\
Figure \ref{Figure2}a shows the schematic conversion of the PSB intensity into an error signal which is then used in a feedback loop to stabilize the electric field experienced by the QD exciton and thus its ZPL emission frequency. The sample used is embedded in a Schottky diode structure which enables the application of an external electric field in order to tune the emission frequency\cite{Warburton2000} through the quantum confined Stark effect. A DC Voltage output of a function generator (FG) is used to tune the QD into resonance with the laser and an additional small-amplitude square wave modulation ($\approx0.5$ mV) is applied at a frequency of $1.5$ kHz. This generates a spectral modulation of the ZPL frequency with an amplitude of $\approx150$ MHz in linear optical frequency (shown as the width of the grey area in Fig. \ref{Figure2}b and c). The spectral modulation correspondingly leads to a detuning-dependent modulation of the PSB mode intensity. The two collected modes are coupled to separate APDs. The PSB mode APD output is connected to the input of a lock-in amplifier in order to generate an error signal, while the ZPL mode is used to characterize the stabilization scheme only and plays no part in the feedback protocol. The demodulated signal at the output of the lock-in amplifier, with a gain of $80$ dB, is shown in Fig. \ref{Figure2}c as a function of laser detuning from the time-averaged resonance frequency. As the modulation is smaller than the linewidth, the error signal is proportional to the derivative of the lineshape. The bandwidth is limited by the integration time needed to build up an error signal larger than the shot noise. For instance, with the count rates shown in Fig. \ref{Figure1} and a detuning of a quarter-linewidth we would need around $6$ ms to acquire an error signal with a signal-to-noise ratio of $1$. Consequently, we work with a gate voltage modulation frequency of $1.5$ kHz and a lock-in time constant of $1-5$ ms. The amplitude of the spectral modulation used here is smaller than the measured absorption linewidth of $610 \pm 20$ MHz. The trion lineshape is consequently broadened to $758 \pm 6$ MHz, while the peak count rate decreases only by $\approx3\%$. Using a higher modulation amplitude increases the size of the error signal obtained and thus also the stabilization bandwidth. However, this also leads to a lower mean count rate. We therefore choose to work with a small modulation amplitude to limit the decrease in peak count rate to a few percent. The error signal is sent to a proportional-integral (PI) controller whose output is fed back as a DC correction to the electric field applied to the QD. \\
We characterize the active stabilization scheme by exciting the QD at $\Omega=\frac{\Gamma}{\sqrt{2}}$ and measuring the noise properties of the RF intensity in the ZPL mode. The black (red) curve in Fig. \ref{Figure3}a shows typical time traces with (without) feedback control and a bin size of $100$ ms. Slow wandering of the QD resonance frequency is apparent in the red curve through temporal fluctuations in the collected count rate which are absent in the stabilized case. Despite stabilizing to a finite root-mean-square detuning from the bare resonance, the mean RF count rate increases from $370$ kHz to $398$ kHz due to the correction of the slow wandering. In order to obtain a more quantitative characterization of the impact of the stabilization scheme, we measure $100$ RF time traces of $30$ s each with a bin size of $200$ $\mathrm{\mu s}$. Figure \ref{Figure3} b shows the average noise spectral density with (without) our stabilization scheme as a black (red) curve obtained by discrete Fourier transform of the time traces. These spectra are normalized such that the integrated power spectral density is equal to the average variance in RF intensity over $30$ s, as follows\cite{Kogan1996,Kuhlmann2013}:
\begin{equation}
N_{\mathrm{QD}} (f) =\frac{(t_{\mathrm{bin}})^{2}}{T} {\left|\mathrm{DFT}\left[\frac{S(i)}{\left<S(i)\right>}\right]\right|}^{2},
\end{equation}   						
where $T$ is the measurement time, $S(i)$ is the number of counts in the ith bin, and $t_{\mathrm{bin}}$ is the bin size. A reference spectrum is taken with a detuned QD and an unsuppressed laser background of similar count rate in order to determine the noise characteristics of the experimental setup itself as well as the shot noise limit. This reference spectrum is then subtracted from the raw spectra in order to measure the noise characteristics of the QD RF only. The red curve shows two components: $1/f$-like noise and an additional Lorentzian decay with a characteristic bandwidth of $20$ Hz due to charge fluctuations in the surrounding matrix\cite{Kuhlmann2013,Stanley2014,Matthiesen2014}. In contrast, the low frequency part of the black curve is flat up to $\approx20$ Hz. This demonstrates the suppression of $1/f$-like noise and slow electric field fluctuations due to charge dynamics in the sample. The noise power of the QD RF is lowered by more than two orders of magnitude at low frequencies, and a smaller improvement can be seen up to frequencies of $\approx1$ kHz. A peak in the noise power at the modulation frequency ($1.5$ kHz) and multiple sidepeaks appear due to the introduction of additional electronic noise into the setup. We note that the introduced fluctuations at the modulation frequency are not critical for applications such as photon interference from independent QDs, as both measurements can be clocked to the same modulation frequency. From the integrated noise power over the frequency range shown, we can extract the contribution of environment-induced noise to the variance in fluorescence intensity over $30$ s. By comparison to the laser-only reference measurement we find that the QD contribution to the overall RF noise reduces from $44$ times the shot noise without stabilization to twice the shot noise with stabilization.\\
We now consider the bandwidth of the stabilization scheme. Comparing the two noise power spectra of Fig \ref{Figure3}.b reveals an improvement up to $\approx 1$ kHz, but this does not correspond directly to a stabilization bandwidth\cite{Prechtel2013}. An improvement of RF noise at frequencies higher than the PID response time is due to a reduction in sensitivity to electric field noise. Specifically, the effect of charge fluctuations on RF intensity noise has a local minimum at zero detuning\cite{Kuhlmann2013,Stanley2014}. Correcting for slow fluctuations in the central frequency of the QD thus leads to a decrease in sensitivity to electric field noise and a subsequent reduction in RF noise power at frequencies above the stabilization bandwidth. For this reason, we consider the intensity autocorrelation instead of the power spectrum, since it allows the identification of specific fluctuation timescales corresponding to different noise sources\cite{Stanley2014,Davanco2014}. The effect of the feedback on the different characteristic correlation decays provides an estimation of the bandwidth of the scheme. Figure 3c shows the intensity autocorrelations extracted from the same dataset as Fig.\ref{Figure3}b. The red points show the average autocorrelation without stabilization and the red curve is a multi-exponential fit comprising $4$ different timescales ($1.2$ ms, $6.8$ ms, $44$ ms, and $752$ ms) with similar amplitudes. The black points show the average autocorrelation of the time traces with the active stabilization, where the oscillations at $1.5$ kHz are due to the gate modulation used. The gray curve shows a multi-exponential fit with timescales fixed by the fit to the unstabilized measurement, and an additional exponentially decaying oscillation representing the gate voltage modulation. Assuming that the underlying charge dynamics remain unaffected, we can quantify the effect of stabilization  by comparing the decay amplitudes occuring on the same timescales in both measurements. Consequently, the two slowest noise sources are fully suppressed, while the amplitudes of the fastest decays of $1.2$ ms and $6.8$ ms are reduced by a factor of $3.1$ and $5.5$, respectively. The fact that the decays are each reduced by different factors signifies that the improvement in intensity noise is not solely due to a reduction in sensitivity and shows unambiguously that the protocol is able to stabilize against frequency fluctuations with a bandwidth higher than some of the charge dynamics in the sample. The $1.5$ kHz oscillations in the autocorrelation decay with a timescale of $5.24$ ms. The amplitude of these oscillations is linked to the error signal acquired whenever the QD is detuned and therefore the decay timescale corresponds to the response time of the stabilization scheme giving a bandwidth of $191 \pm 4$ Hz. This bandwidth is commensurate with the fastest fluctuations associated with electric field noise in self-assembled InAs QDs\cite{Kuhlmann2013,Stanley2014}, therefore the stabilization protocol can partially shield the QD from all electric field noise sources in the environment. Nuclear spin fluctuations occur typically up to $\approx50$ kHz\cite{Kuhlmann2013,Stanley2014} and large improvements in detection efficiency would be required to actively stabilize against this noise source. \\
The aforementioned decrease in RF intensity noise is the consequence of a decrease in ZPL frequency fluctuations due to the active stabilization. To quantify the distribution of ZPL central frequencies we record photon counting histograms of $1$-s long RF time traces with $200$ $\mathrm{\mu s}$ bins, and analyse the effect of the stabilization scheme. These histograms deviate from shot-noise limited Poisson distributions due to fluctuations in the instantaneous detuning of the driving laser from the ZPL transition within the experimental integration time\cite{Matthiesen2014}. By fitting each histogram with a Gaussian detuning distribution of shot-noise limited histograms, we extract both the magnitude of the frequency fluctuations faster than the integration time, as well as the mean laser detuning from resonance. The distribution of mean detunings extracted from a large number of histograms quantifies the variance in the ZPL central frequency and offers an appropriate figure of merit for our stabilization scheme. For the stabilized data we are using a square wave modulation, therefore we assume a detuning distribution consisting of two Gaussian distributions split by the modulation amplitude. The distributions of mean detunings for $781$ time traces with (without) active stabilization are shown in Fig. \ref{Figure3}d and the stabilization scheme clearly shows a significant narrowing of the distribution. Sub-Hz sources of spectral wandering are therefore suppressed fully with this scheme. As our scheme stabilizes to a finite root-mean-square detuning, the peak of the detuning distribution for the stabilized case is at $70$ MHz detuning corresponding to half the modulation amplitude. Slow spectral wandering during the unstabilized measurement leads to a peak in the detuning distribution at a finite detuning of $153$ MHz. From Gaussian fits to the two distributions the FWHM reduces from $154$ MHz to $23$ MHz with the application of active stabilization. We note that the residual spectral wandering of $23$ MHz measured in this way is limited by the standard error in the fitting routine ($23$MHz) and we expect the distribution in central frequencies to be significantly narrower.  \\
In summary, we have developed an undemanding scheme for the stabilization of the ZPL frequency of a solid-state quantum emitter using only the phonon-assisted fluorescence. We have shown that the noise power spectral density of the ZPL fluorescence intensity is reduced up to $1$ kHz with two orders of magnitude improvement in the sub-Hz noise. As well as suppressing the effect of $1/f$-like noise, this stabilization scheme has a sufficient bandwidth to mitigate the effect of the dynamics of several nearby charge traps. The sub-Hz ZPL frequency fluctuations responsible for the largest RF intensity fluctuations are reduced by at least a factor of $7$. A modest improvement in collection and detection efficiency, for instance by embedding the QD in an optimized planar dielectric structure\cite{Lee2011,Chu2014}, would lead to an improvement in stabilization bandwidth permitting the complete suppression of all electric field noise, thereby enabling near-transform-limited emission. This method should greatly improve results from photon coalescence measurements from independent QDs and can be straightforwardly extended to pulsed excitation schemes. Our stabilization scheme does not create additional loss of indistinguishable ZPL photons which is critical to the scalability of QDs as on-demand single photon sources. 
%
\begin{acknowledgments}
We gratefully acknowledge financial support by the University of Cambridge, the European Research Council ERC Consolidator Grant agreement no. 617985, and the EU-FP7 Marie Curie Initial Training Network S3NANO. C.M. gratefully acknowledges Clare College Cambridge for financial support through a Junior Research Fellowship. The authors thank C. Le Gall, R.H.J. Stockill, and J.M. Taylor for fruitful discussions.  
\end{acknowledgments}


\begin{thebibliography}{30}%
\makeatletter
\providecommand \@ifxundefined [1]{%
 \@ifx{#1\undefined}
}%
\providecommand \@ifnum [1]{%
 \ifnum #1\expandafter \@firstoftwo
 \else \expandafter \@secondoftwo
 \fi
}%
\providecommand \@ifx [1]{%
 \ifx #1\expandafter \@firstoftwo
 \else \expandafter \@secondoftwo
 \fi
}%
\providecommand \natexlab [1]{#1}%
\providecommand \enquote  [1]{``#1''}%
\providecommand \bibnamefont  [1]{#1}%
\providecommand \bibfnamefont [1]{#1}%
\providecommand \citenamefont [1]{#1}%
\providecommand \href@noop [0]{\@secondoftwo}%
\providecommand \href [0]{\begingroup \@sanitize@url \@href}%
\providecommand \@href[1]{\@@startlink{#1}\@@href}%
\providecommand \@@href[1]{\endgroup#1\@@endlink}%
\providecommand \@sanitize@url [0]{\catcode `\\12\catcode `\$12\catcode
  `\&12\catcode `\#12\catcode `\^12\catcode `\_12\catcode `\%12\relax}%
\providecommand \@@startlink[1]{}%
\providecommand \@@endlink[0]{}%
\providecommand \url  [0]{\begingroup\@sanitize@url \@url }%
\providecommand \@url [1]{\endgroup\@href {#1}{\urlprefix }}%
\providecommand \urlprefix  [0]{URL }%
\providecommand \Eprint [0]{\href }%
\providecommand \doibase [0]{http://dx.doi.org/}%
\providecommand \selectlanguage [0]{\@gobble}%
\providecommand \bibinfo  [0]{\@secondoftwo}%
\providecommand \bibfield  [0]{\@secondoftwo}%
\providecommand \translation [1]{[#1]}%
\providecommand \BibitemOpen [0]{}%
\providecommand \bibitemStop [0]{}%
\providecommand \bibitemNoStop [0]{.\EOS\space}%
\providecommand \EOS [0]{\spacefactor3000\relax}%
\providecommand \BibitemShut  [1]{\csname bibitem#1\endcsname}%
\let\auto@bib@innerbib\@empty
\bibitem [{\citenamefont {Knill}, \citenamefont {Laflamme},\ and\ \citenamefont
  {Milburn}(2001)}]{Knill2001}%
  \BibitemOpen
  \bibfield  {author} {\bibinfo {author} {\bibfnamefont {E.}~\bibnamefont
  {Knill}}, \bibinfo {author} {\bibfnamefont {R.}~\bibnamefont {Laflamme}}, \
  and\ \bibinfo {author} {\bibfnamefont {G.~J.}\ \bibnamefont {Milburn}},\
  }\href {http://www.nature.com/nature/journal/v409/n6816/abs/409046a0.html}
  {\bibfield  {journal} {\bibinfo  {journal} {Nature (London)}\ }\textbf
  {\bibinfo {volume} {409}},\ \bibinfo {pages} {46} (\bibinfo {year}
  {2001})}\BibitemShut {NoStop}%
\bibitem [{\citenamefont {Kok}\ \emph {et~al.}(2007)\citenamefont {Kok},
  \citenamefont {Munro}, \citenamefont {Nemoto}, \citenamefont {Ralph},
  \citenamefont {Dowling},\ and\ \citenamefont {Milburn}}]{Kok2007}%
  \BibitemOpen
  \bibfield  {author} {\bibinfo {author} {\bibfnamefont {P.}~\bibnamefont
  {Kok}}, \bibinfo {author} {\bibfnamefont {W.~J.}\ \bibnamefont {Munro}},
  \bibinfo {author} {\bibfnamefont {K.}~\bibnamefont {Nemoto}}, \bibinfo
  {author} {\bibfnamefont {T.~C.}\ \bibnamefont {Ralph}}, \bibinfo {author}
  {\bibfnamefont {J.~P.}\ \bibnamefont {Dowling}}, \ and\ \bibinfo {author}
  {\bibfnamefont {G.~J.}\ \bibnamefont {Milburn}},\ }\href {\doibase
  10.1103/RevModPhys.79.135} {\bibfield  {journal} {\bibinfo  {journal} {Rev.
  Mod. Phys.}\ }\textbf {\bibinfo {volume} {79}},\ \bibinfo {pages} {135}
  (\bibinfo {year} {2007})}\BibitemShut {NoStop}%
\bibitem [{\citenamefont {Fr\"ohlich}\ \emph {et~al.}(2013)\citenamefont
  {Fr\"ohlich}, \citenamefont {Dynes}, \citenamefont {Lucamarini},
  \citenamefont {Sharpe}, \citenamefont {Yuan},\ and\ \citenamefont
  {Shields}}]{Froehlich2013}%
  \BibitemOpen
  \bibfield  {author} {\bibinfo {author} {\bibfnamefont {B.}~\bibnamefont
  {Fr\"ohlich}}, \bibinfo {author} {\bibfnamefont {J.~F.}\ \bibnamefont
  {Dynes}}, \bibinfo {author} {\bibfnamefont {M.}~\bibnamefont {Lucamarini}},
  \bibinfo {author} {\bibfnamefont {A.~W.}\ \bibnamefont {Sharpe}}, \bibinfo
  {author} {\bibfnamefont {Z.}~\bibnamefont {Yuan}}, \ and\ \bibinfo {author}
  {\bibfnamefont {A.~J.}\ \bibnamefont {Shields}},\ }\href
  {http://www.nature.com/nature/journal/v501/n7465/full/nature12493.html}
  {\bibfield  {journal} {\bibinfo  {journal} {Nature (London)}\ }\textbf
  {\bibinfo {volume} {501}},\ \bibinfo {pages} {69} (\bibinfo {year}
  {2013})}\BibitemShut {NoStop}%
\bibitem [{\citenamefont {Kimble}(2008)}]{Kimble2008}%
  \BibitemOpen
  \bibfield  {author} {\bibinfo {author} {\bibfnamefont {H.~J.}\ \bibnamefont
  {Kimble}},\ }\href
  {http://www.nature.com/nature/journal/v453/n7198/full/nature07127.html}
  {\bibfield  {journal} {\bibinfo  {journal} {Nature (London)}\ }\textbf
  {\bibinfo {volume} {453}},\ \bibinfo {pages} {1023} (\bibinfo {year}
  {2008})}\BibitemShut {NoStop}%
\bibitem [{\citenamefont {Buckley}, \citenamefont {Rivoire},\ and\
  \citenamefont {Vuckovic}(2012)}]{Buckley2012}%
  \BibitemOpen
  \bibfield  {author} {\bibinfo {author} {\bibfnamefont {S.}~\bibnamefont
  {Buckley}}, \bibinfo {author} {\bibfnamefont {K.}~\bibnamefont {Rivoire}}, \
  and\ \bibinfo {author} {\bibfnamefont {J.}~\bibnamefont {Vuckovic}},\ }\href
  {http://stacks.iop.org/0034-4885/75/i=12/a=126503} {\bibfield  {journal}
  {\bibinfo  {journal} {Reports on Progress in Physics}\ }\textbf {\bibinfo
  {volume} {75}},\ \bibinfo {pages} {126503} (\bibinfo {year}
  {2012})}\BibitemShut {NoStop}%
\bibitem [{\citenamefont {Trotta}\ \emph {et~al.}(2012)\citenamefont {Trotta},
  \citenamefont {Zallo}, \citenamefont {Ortix}, \citenamefont {Atkinson},
  \citenamefont {Plumhof}, \citenamefont {van~den Brink}, \citenamefont
  {Rastelli},\ and\ \citenamefont {Schmidt}}]{Trotta2012}%
  \BibitemOpen
  \bibfield  {author} {\bibinfo {author} {\bibfnamefont {R.}~\bibnamefont
  {Trotta}}, \bibinfo {author} {\bibfnamefont {E.}~\bibnamefont {Zallo}},
  \bibinfo {author} {\bibfnamefont {C.}~\bibnamefont {Ortix}}, \bibinfo
  {author} {\bibfnamefont {P.}~\bibnamefont {Atkinson}}, \bibinfo {author}
  {\bibfnamefont {J.~D.}\ \bibnamefont {Plumhof}}, \bibinfo {author}
  {\bibfnamefont {J.}~\bibnamefont {van~den Brink}}, \bibinfo {author}
  {\bibfnamefont {A.}~\bibnamefont {Rastelli}}, \ and\ \bibinfo {author}
  {\bibfnamefont {O.~G.}\ \bibnamefont {Schmidt}},\ }\href {\doibase
  10.1103/PhysRevLett.109.147401} {\bibfield  {journal} {\bibinfo  {journal}
  {Phys. Rev. Lett.}\ }\textbf {\bibinfo {volume} {109}},\ \bibinfo {pages}
  {147401} (\bibinfo {year} {2012})}\BibitemShut {NoStop}%
\bibitem [{\citenamefont {Bennett}\ \emph {et~al.}(2005)\citenamefont
  {Bennett}, \citenamefont {Unitt}, \citenamefont {See}, \citenamefont
  {Shields}, \citenamefont {Atkinson}, \citenamefont {Cooper},\ and\
  \citenamefont {Ritchie}}]{Bennett2005}%
  \BibitemOpen
  \bibfield  {author} {\bibinfo {author} {\bibfnamefont {A.~J.}\ \bibnamefont
  {Bennett}}, \bibinfo {author} {\bibfnamefont {D.~C.}\ \bibnamefont {Unitt}},
  \bibinfo {author} {\bibfnamefont {P.}~\bibnamefont {See}}, \bibinfo {author}
  {\bibfnamefont {A.~J.}\ \bibnamefont {Shields}}, \bibinfo {author}
  {\bibfnamefont {P.}~\bibnamefont {Atkinson}}, \bibinfo {author}
  {\bibfnamefont {K.}~\bibnamefont {Cooper}}, \ and\ \bibinfo {author}
  {\bibfnamefont {D.~A.}\ \bibnamefont {Ritchie}},\ }\href {\doibase
  10.1103/PhysRevB.72.033316} {\bibfield  {journal} {\bibinfo  {journal} {Phys.
  Rev. B}\ }\textbf {\bibinfo {volume} {72}},\ \bibinfo {pages} {033316}
  (\bibinfo {year} {2005})}\BibitemShut {NoStop}%
\bibitem [{\citenamefont {Hargart}\ \emph {et~al.}(2013)\citenamefont
  {Hargart}, \citenamefont {Kessler}, \citenamefont {Schwarzbäck},
  \citenamefont {Koroknay}, \citenamefont {Weidenfeld}, \citenamefont
  {Jetter},\ and\ \citenamefont {Michler}}]{Hargart2013}%
  \BibitemOpen
  \bibfield  {author} {\bibinfo {author} {\bibfnamefont {F.}~\bibnamefont
  {Hargart}}, \bibinfo {author} {\bibfnamefont {C.~A.}\ \bibnamefont
  {Kessler}}, \bibinfo {author} {\bibfnamefont {T.}~\bibnamefont
  {Schwarzbäck}}, \bibinfo {author} {\bibfnamefont {E.}~\bibnamefont
  {Koroknay}}, \bibinfo {author} {\bibfnamefont {S.}~\bibnamefont
  {Weidenfeld}}, \bibinfo {author} {\bibfnamefont {M.}~\bibnamefont {Jetter}},
  \ and\ \bibinfo {author} {\bibfnamefont {P.}~\bibnamefont {Michler}},\ }\href
  {\doibase http://dx.doi.org/10.1063/1.4774392} {\bibfield  {journal}
  {\bibinfo  {journal} {Applied Physics Letters}\ }\textbf {\bibinfo {volume}
  {102}},\ \bibinfo {eid} {011126} (\bibinfo {year} {2013})}\BibitemShut
  {NoStop}%
\bibitem [{\citenamefont {Lodahl}, \citenamefont {Mahmoodian},\ and\
  \citenamefont {Stobbe}(2013)}]{Lodahl2013}%
  \BibitemOpen
  \bibfield  {author} {\bibinfo {author} {\bibfnamefont {P.}~\bibnamefont
  {Lodahl}}, \bibinfo {author} {\bibfnamefont {S.}~\bibnamefont {Mahmoodian}},
  \ and\ \bibinfo {author} {\bibfnamefont {S.}~\bibnamefont {Stobbe}},\ }\href
  {http://arxiv.org/abs/1312.1079} {\bibfield  {journal} {\bibinfo  {journal}
  {arXiv preprint}\ }\textbf {\bibinfo {volume} {arXiv:1312.1079v1}} (\bibinfo
  {year} {2013})}\BibitemShut {NoStop}%
\bibitem [{\citenamefont {Kuhlmann}\ \emph {et~al.}(2013)\citenamefont
  {Kuhlmann}, \citenamefont {Houel}, \citenamefont {Ludwig}, \citenamefont
  {Greuter}, \citenamefont {Reuter}, \citenamefont {Wieck}, \citenamefont
  {Poggio},\ and\ \citenamefont {Warburton}}]{Kuhlmann2013}%
  \BibitemOpen
  \bibfield  {author} {\bibinfo {author} {\bibfnamefont {A.~V.}\ \bibnamefont
  {Kuhlmann}}, \bibinfo {author} {\bibfnamefont {J.}~\bibnamefont {Houel}},
  \bibinfo {author} {\bibfnamefont {A.}~\bibnamefont {Ludwig}}, \bibinfo
  {author} {\bibfnamefont {L.}~\bibnamefont {Greuter}}, \bibinfo {author}
  {\bibfnamefont {D.}~\bibnamefont {Reuter}}, \bibinfo {author} {\bibfnamefont
  {A.~D.}\ \bibnamefont {Wieck}}, \bibinfo {author} {\bibfnamefont
  {M.}~\bibnamefont {Poggio}}, \ and\ \bibinfo {author} {\bibfnamefont {R.~J.}\
  \bibnamefont {Warburton}},\ }\href
  {http://www.nature.com/nphys/journal/v9/n9/full/nphys2688.html} {\bibfield
  {journal} {\bibinfo  {journal} {Nature Physics}\ }\textbf {\bibinfo {volume}
  {9}},\ \bibinfo {pages} {570} (\bibinfo {year} {2013})}\BibitemShut {NoStop}%
\bibitem [{\citenamefont {Stanley}\ \emph {et~al.}(2014)\citenamefont
  {Stanley}, \citenamefont {Matthiesen}, \citenamefont {Hansom}, \citenamefont
  {Le~Gall}, \citenamefont {Schulte}, \citenamefont {Clarke},\ and\
  \citenamefont {Atat\"ure}}]{Stanley2014}%
  \BibitemOpen
  \bibfield  {author} {\bibinfo {author} {\bibfnamefont {M.~J.}\ \bibnamefont
  {Stanley}}, \bibinfo {author} {\bibfnamefont {C.}~\bibnamefont {Matthiesen}},
  \bibinfo {author} {\bibfnamefont {J.}~\bibnamefont {Hansom}}, \bibinfo
  {author} {\bibfnamefont {C.}~\bibnamefont {Le~Gall}}, \bibinfo {author}
  {\bibfnamefont {C.~H.~H.}\ \bibnamefont {Schulte}}, \bibinfo {author}
  {\bibfnamefont {E.}~\bibnamefont {Clarke}}, \ and\ \bibinfo {author}
  {\bibfnamefont {M.}~\bibnamefont {Atat\"ure}},\ }\href
  {http://arxiv.org/abs/1408.6437} {\bibfield  {journal} {\bibinfo  {journal}
  {arXiv preprint}\ }\textbf {\bibinfo {volume} {arXiv:1408.6437v1}} (\bibinfo
  {year} {2014})}\BibitemShut {NoStop}%
\bibitem [{\citenamefont {Matthiesen}\ \emph {et~al.}(2014)\citenamefont
  {Matthiesen}, \citenamefont {Stanley}, \citenamefont {Hugues}, \citenamefont
  {Clarke},\ and\ \citenamefont {Atat\"ure}}]{Matthiesen2014}%
  \BibitemOpen
  \bibfield  {author} {\bibinfo {author} {\bibfnamefont {C.}~\bibnamefont
  {Matthiesen}}, \bibinfo {author} {\bibfnamefont {M.~J.}\ \bibnamefont
  {Stanley}}, \bibinfo {author} {\bibfnamefont {M.}~\bibnamefont {Hugues}},
  \bibinfo {author} {\bibfnamefont {E.}~\bibnamefont {Clarke}}, \ and\ \bibinfo
  {author} {\bibfnamefont {M.}~\bibnamefont {Atat\"ure}},\ }\href {\doibase
  10.1038/srep04911} {\bibfield  {journal} {\bibinfo  {journal} {Scientific
  Reports}\ }\textbf {\bibinfo {volume} {4}},\ \bibinfo {pages} {4911}
  (\bibinfo {year} {2014})}\BibitemShut {NoStop}%
\bibitem [{\citenamefont {Varnava}, \citenamefont {Browne},\ and\ \citenamefont
  {Rudolph}(2008)}]{Varnava2008}%
  \BibitemOpen
  \bibfield  {author} {\bibinfo {author} {\bibfnamefont {M.}~\bibnamefont
  {Varnava}}, \bibinfo {author} {\bibfnamefont {D.~E.}\ \bibnamefont {Browne}},
  \ and\ \bibinfo {author} {\bibfnamefont {T.}~\bibnamefont {Rudolph}},\ }\href
  {\doibase 10.1103/PhysRevLett.100.060502} {\bibfield  {journal} {\bibinfo
  {journal} {Phys. Rev. Lett.}\ }\textbf {\bibinfo {volume} {100}},\ \bibinfo
  {pages} {060502} (\bibinfo {year} {2008})}\BibitemShut {NoStop}%
\bibitem [{\citenamefont {Gong}\ \emph {et~al.}(2010)\citenamefont {Gong},
  \citenamefont {Zou}, \citenamefont {Ralph}, \citenamefont {Zhu},\ and\
  \citenamefont {Guo}}]{Gong2010}%
  \BibitemOpen
  \bibfield  {author} {\bibinfo {author} {\bibfnamefont {Y.-X.}\ \bibnamefont
  {Gong}}, \bibinfo {author} {\bibfnamefont {X.-B.}\ \bibnamefont {Zou}},
  \bibinfo {author} {\bibfnamefont {T.~C.}\ \bibnamefont {Ralph}}, \bibinfo
  {author} {\bibfnamefont {S.-N.}\ \bibnamefont {Zhu}}, \ and\ \bibinfo
  {author} {\bibfnamefont {G.-C.}\ \bibnamefont {Guo}},\ }\href {\doibase
  10.1103/PhysRevA.81.052303} {\bibfield  {journal} {\bibinfo  {journal} {Phys.
  Rev. A}\ }\textbf {\bibinfo {volume} {81}},\ \bibinfo {pages} {052303}
  (\bibinfo {year} {2010})}\BibitemShut {NoStop}%
\bibitem [{\citenamefont {Acosta}\ \emph {et~al.}(2012)\citenamefont {Acosta},
  \citenamefont {Santori}, \citenamefont {Faraon}, \citenamefont {Huang},
  \citenamefont {Fu}, \citenamefont {Stacey}, \citenamefont {Simpson},
  \citenamefont {Ganesan}, \citenamefont {Tomljenovic-Hanic}, \citenamefont
  {Greentree}, \citenamefont {Prawer},\ and\ \citenamefont
  {Beausoleil}}]{Acosta2012}%
  \BibitemOpen
  \bibfield  {author} {\bibinfo {author} {\bibfnamefont {V.~M.}\ \bibnamefont
  {Acosta}}, \bibinfo {author} {\bibfnamefont {C.}~\bibnamefont {Santori}},
  \bibinfo {author} {\bibfnamefont {A.}~\bibnamefont {Faraon}}, \bibinfo
  {author} {\bibfnamefont {Z.}~\bibnamefont {Huang}}, \bibinfo {author}
  {\bibfnamefont {K.-M.~C.}\ \bibnamefont {Fu}}, \bibinfo {author}
  {\bibfnamefont {A.}~\bibnamefont {Stacey}}, \bibinfo {author} {\bibfnamefont
  {D.~A.}\ \bibnamefont {Simpson}}, \bibinfo {author} {\bibfnamefont
  {K.}~\bibnamefont {Ganesan}}, \bibinfo {author} {\bibfnamefont
  {S.}~\bibnamefont {Tomljenovic-Hanic}}, \bibinfo {author} {\bibfnamefont
  {A.~D.}\ \bibnamefont {Greentree}}, \bibinfo {author} {\bibfnamefont
  {S.}~\bibnamefont {Prawer}}, \ and\ \bibinfo {author} {\bibfnamefont {R.~G.}\
  \bibnamefont {Beausoleil}},\ }\href {\doibase 10.1103/PhysRevLett.108.206401}
  {\bibfield  {journal} {\bibinfo  {journal} {Phys. Rev. Lett.}\ }\textbf
  {\bibinfo {volume} {108}},\ \bibinfo {pages} {206401} (\bibinfo {year}
  {2012})}\BibitemShut {NoStop}%
\bibitem [{\citenamefont {Akopian}\ \emph {et~al.}(2013)\citenamefont
  {Akopian}, \citenamefont {Trotta}, \citenamefont {Zallo}, \citenamefont
  {Kumar}, \citenamefont {Atkinson}, \citenamefont {Rastelli}, \citenamefont
  {Schmidt},\ and\ \citenamefont {Zwiller}}]{Akopian2013}%
  \BibitemOpen
  \bibfield  {author} {\bibinfo {author} {\bibfnamefont {N.}~\bibnamefont
  {Akopian}}, \bibinfo {author} {\bibfnamefont {R.}~\bibnamefont {Trotta}},
  \bibinfo {author} {\bibfnamefont {E.}~\bibnamefont {Zallo}}, \bibinfo
  {author} {\bibfnamefont {S.}~\bibnamefont {Kumar}}, \bibinfo {author}
  {\bibfnamefont {P.}~\bibnamefont {Atkinson}}, \bibinfo {author}
  {\bibfnamefont {A.}~\bibnamefont {Rastelli}}, \bibinfo {author}
  {\bibfnamefont {O.~G.}\ \bibnamefont {Schmidt}}, \ and\ \bibinfo {author}
  {\bibfnamefont {V.}~\bibnamefont {Zwiller}},\ }\href
  {http://arxiv.org/abs/1302.2005} {\bibfield  {journal} {\bibinfo  {journal}
  {arXiv preprint}\ }\textbf {\bibinfo {volume} {arXiv:1302.2005v1}} (\bibinfo
  {year} {2013})}\BibitemShut {NoStop}%
\bibitem [{\citenamefont {Prechtel}\ \emph {et~al.}(2013)\citenamefont
  {Prechtel}, \citenamefont {Kuhlmann}, \citenamefont {Houel}, \citenamefont
  {Greuter}, \citenamefont {Ludwig}, \citenamefont {Reuter}, \citenamefont
  {Wieck},\ and\ \citenamefont {Warburton}}]{Prechtel2013}%
  \BibitemOpen
  \bibfield  {author} {\bibinfo {author} {\bibfnamefont {J.~H.}\ \bibnamefont
  {Prechtel}}, \bibinfo {author} {\bibfnamefont {A.~V.}\ \bibnamefont
  {Kuhlmann}}, \bibinfo {author} {\bibfnamefont {J.}~\bibnamefont {Houel}},
  \bibinfo {author} {\bibfnamefont {L.}~\bibnamefont {Greuter}}, \bibinfo
  {author} {\bibfnamefont {A.}~\bibnamefont {Ludwig}}, \bibinfo {author}
  {\bibfnamefont {D.}~\bibnamefont {Reuter}}, \bibinfo {author} {\bibfnamefont
  {A.~D.}\ \bibnamefont {Wieck}}, \ and\ \bibinfo {author} {\bibfnamefont
  {R.~J.}\ \bibnamefont {Warburton}},\ }\href {\doibase
  10.1103/PhysRevX.3.041006} {\bibfield  {journal} {\bibinfo  {journal} {Phys.
  Rev. X}\ }\textbf {\bibinfo {volume} {3}},\ \bibinfo {pages} {041006}
  (\bibinfo {year} {2013})}\BibitemShut {NoStop}%
\bibitem [{\citenamefont {Besombes}\ \emph {et~al.}(2001)\citenamefont
  {Besombes}, \citenamefont {Kheng}, \citenamefont {Marsal},\ and\
  \citenamefont {Mariette}}]{Besombes2001}%
  \BibitemOpen
  \bibfield  {author} {\bibinfo {author} {\bibfnamefont {L.}~\bibnamefont
  {Besombes}}, \bibinfo {author} {\bibfnamefont {K.}~\bibnamefont {Kheng}},
  \bibinfo {author} {\bibfnamefont {L.}~\bibnamefont {Marsal}}, \ and\ \bibinfo
  {author} {\bibfnamefont {H.}~\bibnamefont {Mariette}},\ }\href {\doibase
  10.1103/PhysRevB.63.155307} {\bibfield  {journal} {\bibinfo  {journal} {Phys.
  Rev. B}\ }\textbf {\bibinfo {volume} {63}},\ \bibinfo {pages} {155307}
  (\bibinfo {year} {2001})}\BibitemShut {NoStop}%
\bibitem [{\citenamefont {Favero}\ \emph {et~al.}(2003)\citenamefont {Favero},
  \citenamefont {Cassabois}, \citenamefont {Ferreira}, \citenamefont {Darson},
  \citenamefont {Voisin}, \citenamefont {Tignon}, \citenamefont {Delalande},
  \citenamefont {Bastard}, \citenamefont {Roussignol},\ and\ \citenamefont
  {G\'erard}}]{Favero2003}%
  \BibitemOpen
  \bibfield  {author} {\bibinfo {author} {\bibfnamefont {I.}~\bibnamefont
  {Favero}}, \bibinfo {author} {\bibfnamefont {G.}~\bibnamefont {Cassabois}},
  \bibinfo {author} {\bibfnamefont {R.}~\bibnamefont {Ferreira}}, \bibinfo
  {author} {\bibfnamefont {D.}~\bibnamefont {Darson}}, \bibinfo {author}
  {\bibfnamefont {C.}~\bibnamefont {Voisin}}, \bibinfo {author} {\bibfnamefont
  {J.}~\bibnamefont {Tignon}}, \bibinfo {author} {\bibfnamefont
  {C.}~\bibnamefont {Delalande}}, \bibinfo {author} {\bibfnamefont
  {G.}~\bibnamefont {Bastard}}, \bibinfo {author} {\bibfnamefont
  {P.}~\bibnamefont {Roussignol}}, \ and\ \bibinfo {author} {\bibfnamefont
  {J.~M.}\ \bibnamefont {G\'erard}},\ }\href {\doibase
  10.1103/PhysRevB.68.233301} {\bibfield  {journal} {\bibinfo  {journal} {Phys.
  Rev. B}\ }\textbf {\bibinfo {volume} {68}},\ \bibinfo {pages} {233301}
  (\bibinfo {year} {2003})}\BibitemShut {NoStop}%
\bibitem [{\citenamefont {Krummheuer}, \citenamefont {Axt},\ and\ \citenamefont
  {Kuhn}(2002)}]{Krummheuer2002}%
  \BibitemOpen
  \bibfield  {author} {\bibinfo {author} {\bibfnamefont {B.}~\bibnamefont
  {Krummheuer}}, \bibinfo {author} {\bibfnamefont {V.~M.}\ \bibnamefont {Axt}},
  \ and\ \bibinfo {author} {\bibfnamefont {T.}~\bibnamefont {Kuhn}},\ }\href
  {\doibase 10.1103/PhysRevB.65.195313} {\bibfield  {journal} {\bibinfo
  {journal} {Phys. Rev. B}\ }\textbf {\bibinfo {volume} {65}},\ \bibinfo
  {pages} {195313} (\bibinfo {year} {2002})}\BibitemShut {NoStop}%
\bibitem [{\citenamefont {Matthiesen}\ \emph {et~al.}(2013)\citenamefont
  {Matthiesen}, \citenamefont {Geller}, \citenamefont {Schulte}, \citenamefont
  {Gall}, \citenamefont {Hansom}, \citenamefont {Li}, \citenamefont {Hugues},
  \citenamefont {Clarke},\ and\ \citenamefont {Atat\"ure}}]{Matthiesen2013}%
  \BibitemOpen
  \bibfield  {author} {\bibinfo {author} {\bibfnamefont {C.}~\bibnamefont
  {Matthiesen}}, \bibinfo {author} {\bibfnamefont {M.}~\bibnamefont {Geller}},
  \bibinfo {author} {\bibfnamefont {C.~H.~H.}\ \bibnamefont {Schulte}},
  \bibinfo {author} {\bibfnamefont {C.~L.}\ \bibnamefont {Gall}}, \bibinfo
  {author} {\bibfnamefont {J.}~\bibnamefont {Hansom}}, \bibinfo {author}
  {\bibfnamefont {Z.}~\bibnamefont {Li}}, \bibinfo {author} {\bibfnamefont
  {M.}~\bibnamefont {Hugues}}, \bibinfo {author} {\bibfnamefont
  {E.}~\bibnamefont {Clarke}}, \ and\ \bibinfo {author} {\bibfnamefont
  {M.}~\bibnamefont {Atat\"ure}},\ }\href {\doibase 10.1038/ncomms2601}
  {\bibfield  {journal} {\bibinfo  {journal} {Nature Communications}\ }\textbf
  {\bibinfo {volume} {4}},\ \bibinfo {pages} {1600} (\bibinfo {year}
  {2013})}\BibitemShut {NoStop}%
\bibitem [{\citenamefont {He}\ \emph {et~al.}(2013)\citenamefont {He},
  \citenamefont {He}, \citenamefont {Wei}, \citenamefont {Wu}, \citenamefont
  {Atat\"ure}, \citenamefont {Schneider}, \citenamefont {H\"ofling},
  \citenamefont {Kamp}, \citenamefont {Lu},\ and\ \citenamefont
  {Pan}}]{He2013}%
  \BibitemOpen
  \bibfield  {author} {\bibinfo {author} {\bibfnamefont {Y.-M.}\ \bibnamefont
  {He}}, \bibinfo {author} {\bibfnamefont {Y.}~\bibnamefont {He}}, \bibinfo
  {author} {\bibfnamefont {Y.-J.}\ \bibnamefont {Wei}}, \bibinfo {author}
  {\bibfnamefont {D.}~\bibnamefont {Wu}}, \bibinfo {author} {\bibfnamefont
  {M.}~\bibnamefont {Atat\"ure}}, \bibinfo {author} {\bibfnamefont
  {C.}~\bibnamefont {Schneider}}, \bibinfo {author} {\bibfnamefont
  {S.}~\bibnamefont {H\"ofling}}, \bibinfo {author} {\bibfnamefont
  {M.}~\bibnamefont {Kamp}}, \bibinfo {author} {\bibfnamefont {C.-Y.}\
  \bibnamefont {Lu}}, \ and\ \bibinfo {author} {\bibfnamefont {J.-W.}\
  \bibnamefont {Pan}},\ }\href {\doibase 10.1038/nnano.2012.262} {\bibfield
  {journal} {\bibinfo  {journal} {Nature Nanotechnology}\ }\textbf {\bibinfo
  {volume} {8}},\ \bibinfo {pages} {213} (\bibinfo {year} {2013})}\BibitemShut
  {NoStop}%
\bibitem [{\citenamefont {Gao}\ \emph {et~al.}(2013)\citenamefont {Gao},
  \citenamefont {Fallahi}, \citenamefont {Togan}, \citenamefont {Delteil},
  \citenamefont {Chin}, \citenamefont {Miguel-Sanchez},\ and\ \citenamefont
  {Imamoglu}}]{Gao2013}%
  \BibitemOpen
  \bibfield  {author} {\bibinfo {author} {\bibfnamefont {W.}~\bibnamefont
  {Gao}}, \bibinfo {author} {\bibfnamefont {P.}~\bibnamefont {Fallahi}},
  \bibinfo {author} {\bibfnamefont {E.}~\bibnamefont {Togan}}, \bibinfo
  {author} {\bibfnamefont {A.}~\bibnamefont {Delteil}}, \bibinfo {author}
  {\bibfnamefont {Y.}~\bibnamefont {Chin}}, \bibinfo {author} {\bibfnamefont
  {J.}~\bibnamefont {Miguel-Sanchez}}, \ and\ \bibinfo {author} {\bibfnamefont
  {A.}~\bibnamefont {Imamoglu}},\ }\href {\doibase 10.1038/ncomms3744}
  {\bibfield  {journal} {\bibinfo  {journal} {Nature Communications}\ }\textbf
  {\bibinfo {volume} {4}},\ \bibinfo {pages} {2744} (\bibinfo {year}
  {2013})}\BibitemShut {NoStop}%
\bibitem [{\citenamefont {Gazzano}\ \emph {et~al.}(2013)\citenamefont
  {Gazzano}, \citenamefont {de~Vasconcellos}, \citenamefont {Arnold},
  \citenamefont {Nowak}, \citenamefont {Galopin}, \citenamefont {Sagnes},
  \citenamefont {Lanco}, \citenamefont {Lema\^itre},\ and\ \citenamefont
  {Senellart}}]{Gazzano2013}%
  \BibitemOpen
  \bibfield  {author} {\bibinfo {author} {\bibfnamefont {O.}~\bibnamefont
  {Gazzano}}, \bibinfo {author} {\bibfnamefont {S.~M.}\ \bibnamefont
  {de~Vasconcellos}}, \bibinfo {author} {\bibfnamefont {C.}~\bibnamefont
  {Arnold}}, \bibinfo {author} {\bibfnamefont {A.}~\bibnamefont {Nowak}},
  \bibinfo {author} {\bibfnamefont {E.}~\bibnamefont {Galopin}}, \bibinfo
  {author} {\bibfnamefont {I.}~\bibnamefont {Sagnes}}, \bibinfo {author}
  {\bibfnamefont {L.}~\bibnamefont {Lanco}}, \bibinfo {author} {\bibfnamefont
  {A.}~\bibnamefont {Lema\^itre}}, \ and\ \bibinfo {author} {\bibfnamefont
  {P.}~\bibnamefont {Senellart}},\ }\href
  {http://www.nature.com/nature/journal/v405/n6789/abs/405926a0.html}
  {\bibfield  {journal} {\bibinfo  {journal} {Nature Communications}\ }\textbf
  {\bibinfo {volume} {4}},\ \bibinfo {pages} {1425} (\bibinfo {year}
  {2013})}\BibitemShut {NoStop}%
\bibitem [{\citenamefont {Matthiesen}, \citenamefont {Vamivakas},\ and\
  \citenamefont {Atat\"ure}(2012)}]{Matthiesen2012}%
  \BibitemOpen
  \bibfield  {author} {\bibinfo {author} {\bibfnamefont {C.}~\bibnamefont
  {Matthiesen}}, \bibinfo {author} {\bibfnamefont {A.~N.}\ \bibnamefont
  {Vamivakas}}, \ and\ \bibinfo {author} {\bibfnamefont {M.}~\bibnamefont
  {Atat\"ure}},\ }\href {\doibase 10.1103/PhysRevLett.108.093602} {\bibfield
  {journal} {\bibinfo  {journal} {Phys. Rev. Lett.}\ }\textbf {\bibinfo
  {volume} {108}},\ \bibinfo {pages} {093602} (\bibinfo {year}
  {2012})}\BibitemShut {NoStop}%
\bibitem [{\citenamefont {Warburton}\ \emph {et~al.}(2000)\citenamefont
  {Warburton}, \citenamefont {Sch\"aflein}, \citenamefont {Haft}, \citenamefont
  {Bickel}, \citenamefont {Lorke}, \citenamefont {Karrai}, \citenamefont
  {Garcia}, \citenamefont {Schoenfeld},\ and\ \citenamefont
  {Petroff}}]{Warburton2000}%
  \BibitemOpen
  \bibfield  {author} {\bibinfo {author} {\bibfnamefont {R.~J.}\ \bibnamefont
  {Warburton}}, \bibinfo {author} {\bibfnamefont {C.}~\bibnamefont
  {Sch\"aflein}}, \bibinfo {author} {\bibfnamefont {D.}~\bibnamefont {Haft}},
  \bibinfo {author} {\bibfnamefont {F.}~\bibnamefont {Bickel}}, \bibinfo
  {author} {\bibfnamefont {A.}~\bibnamefont {Lorke}}, \bibinfo {author}
  {\bibfnamefont {K.}~\bibnamefont {Karrai}}, \bibinfo {author} {\bibfnamefont
  {J.~M.}\ \bibnamefont {Garcia}}, \bibinfo {author} {\bibfnamefont
  {W.}~\bibnamefont {Schoenfeld}}, \ and\ \bibinfo {author} {\bibfnamefont
  {P.~M.}\ \bibnamefont {Petroff}},\ }\href
  {http://www.nature.com/nature/journal/v405/n6789/abs/405926a0.html}
  {\bibfield  {journal} {\bibinfo  {journal} {Nature (London)}\ }\textbf
  {\bibinfo {volume} {405}},\ \bibinfo {pages} {926} (\bibinfo {year}
  {2000})}\BibitemShut {NoStop}%
\bibitem [{\citenamefont {Kogan}(1996)}]{Kogan1996}%
  \BibitemOpen
  \bibfield  {author} {\bibinfo {author} {\bibfnamefont {S.}~\bibnamefont
  {Kogan}},\ }\href
  {http://ebooks.cambridge.org/ebook.jsf?bid=CBO9780511551666} {\emph {\bibinfo
  {title} {Electronic Noise and Fluctuations in Solids}}}\ (\bibinfo
  {publisher} {Cambridge University Press},\ \bibinfo {year}
  {1996})\BibitemShut {NoStop}%
\bibitem [{\citenamefont {Davan\ifmmode~\mbox{\c{c}}\else \c{c}\fi{}o}\ \emph
  {et~al.}(2014)\citenamefont {Davan\ifmmode~\mbox{\c{c}}\else \c{c}\fi{}o},
  \citenamefont {Hellberg}, \citenamefont {Ates}, \citenamefont {Badolato},\
  and\ \citenamefont {Srinivasan}}]{Davanco2014}%
  \BibitemOpen
  \bibfield  {author} {\bibinfo {author} {\bibfnamefont {M.}~\bibnamefont
  {Davan\ifmmode~\mbox{\c{c}}\else \c{c}\fi{}o}}, \bibinfo {author}
  {\bibfnamefont {C.~S.}\ \bibnamefont {Hellberg}}, \bibinfo {author}
  {\bibfnamefont {S.}~\bibnamefont {Ates}}, \bibinfo {author} {\bibfnamefont
  {A.}~\bibnamefont {Badolato}}, \ and\ \bibinfo {author} {\bibfnamefont
  {K.}~\bibnamefont {Srinivasan}},\ }\href {\doibase
  10.1103/PhysRevB.89.161303} {\bibfield  {journal} {\bibinfo  {journal} {Phys.
  Rev. B}\ }\textbf {\bibinfo {volume} {89}},\ \bibinfo {pages} {161303}
  (\bibinfo {year} {2014})}\BibitemShut {NoStop}%
\bibitem [{\citenamefont {Lee}\ \emph {et~al.}(2011)\citenamefont {Lee},
  \citenamefont {Chen}, \citenamefont {Eghlidi}, \citenamefont {Kukura},
  \citenamefont {Lettow}, \citenamefont {Renn}, \citenamefont {Sandoghdar},\
  and\ \citenamefont {G\"otzinger}}]{Lee2011}%
  \BibitemOpen
  \bibfield  {author} {\bibinfo {author} {\bibfnamefont {K.~G.}\ \bibnamefont
  {Lee}}, \bibinfo {author} {\bibfnamefont {X.~W.}\ \bibnamefont {Chen}},
  \bibinfo {author} {\bibfnamefont {H.}~\bibnamefont {Eghlidi}}, \bibinfo
  {author} {\bibfnamefont {P.}~\bibnamefont {Kukura}}, \bibinfo {author}
  {\bibfnamefont {R.}~\bibnamefont {Lettow}}, \bibinfo {author} {\bibfnamefont
  {A.}~\bibnamefont {Renn}}, \bibinfo {author} {\bibfnamefont {V.}~\bibnamefont
  {Sandoghdar}}, \ and\ \bibinfo {author} {\bibfnamefont {S.}~\bibnamefont
  {G\"otzinger}},\ }\href
  {http://www.nature.com/nphoton/journal/v5/n3/full/nphoton.2010.312.html}
  {\bibfield  {journal} {\bibinfo  {journal} {Nature Photonics}\ }\textbf
  {\bibinfo {volume} {5}},\ \bibinfo {pages} {166} (\bibinfo {year}
  {2011})}\BibitemShut {NoStop}%
\bibitem [{\citenamefont {Chu}\ \emph {et~al.}(2014)\citenamefont {Chu},
  \citenamefont {Brenner}, \citenamefont {Chen}, \citenamefont {Ghosh},
  \citenamefont {Hollingsworth}, \citenamefont {Sandoghdar},\ and\
  \citenamefont {G\"otzinger}}]{Chu2014}%
  \BibitemOpen
  \bibfield  {author} {\bibinfo {author} {\bibfnamefont {X.-L.}\ \bibnamefont
  {Chu}}, \bibinfo {author} {\bibfnamefont {T.}~\bibnamefont {Brenner}},
  \bibinfo {author} {\bibfnamefont {X.-W.}\ \bibnamefont {Chen}}, \bibinfo
  {author} {\bibfnamefont {Y.}~\bibnamefont {Ghosh}}, \bibinfo {author}
  {\bibfnamefont {J.~A.}\ \bibnamefont {Hollingsworth}}, \bibinfo {author}
  {\bibfnamefont {V.}~\bibnamefont {Sandoghdar}}, \ and\ \bibinfo {author}
  {\bibfnamefont {S.}~\bibnamefont {G\"otzinger}},\ }\href
  {http://arxiv.org/abs/1406.0626} {\bibfield  {journal} {\bibinfo  {journal}
  {arXiv preprint}\ }\textbf {\bibinfo {volume} {arXiv:1406.0626v1}} (\bibinfo
  {year} {2014})}\BibitemShut {NoStop}%
\end{thebibliography}

%

\end{document}